\documentclass{aa}  
\usepackage{graphicx}
\usepackage{txfonts}
\usepackage{hyperref}

\newcommand{\REV}[1]{{#1}}

\begin{document}

   \title{Beyond the largest Lyapunov exponent: Entropy-based diagnostics of chaos in Hénon-Heiles and $N$-body dynamics}
    \titlerunning {Entropy-based diagnostic of chaos}
   \author{    
        Alessandro Alberto Trani\inst{1,2} 
        \and
        Pierfrancesco Di Cintio
        \inst{3,4,5} \email{pierfrancesco.dicintio@cnr.it}      
        \and
        Michele Ginolfi
        \inst{3,6} \email{michele.ginolfi@unifi.it}
        }
  \institute{
        Department of Astronomy, University of Concepci\'on, Avenida Esteban Iturra s/n
Casilla 160-C Concepci\'on, Chile\and
    National Institute for Nuclear Physics – INFN, Sezione di Trieste, I-34127, Trieste, Italy\\   
\and
    National Institute of Astrophysics - Arcetri Astrophysical Observatory (INAF-OAA), Piazzale E.\ Fermi 5, I-50125 Firenze, Italy\and
 National Council of Research - Institute of Complex Systems, Via Madonna del piano 10, I-50019 Sesto Fiorentino, Italy
        \and
        National Institute of Nuclear Physics (INFN) -  Florence unit, via G. Sansone 1, I-50019 Sesto Fiorentino, Italy\\
\and
 University of Florence, Department of Physics and Astronomy - Astrophysics and Space Science section, Piazzale E.\ Fermi 5, I-50125 Firenze, Italy \\
    }
   \date{Received M DD, YYYY; accepted M DD, YYYY}
  \abstract
   {The largest Lyapunov exponent is widely used to diagnose chaos in gravitational dynamics; however, in mixed phase spaces and finite-$N$ systems, it does not always provide a complete description of the orbital complexity and phase-space transport. Entropy-based diagnostics could offer a complementary perspective.}
   {
   We investigate whether trajectory-based information entropy can provide a useful diagnostic of chaos in gravitational systems and how it relates to the largest Lyapunov exponent as a function of orbital energy and of the number of degrees of freedom.}
   {
   We computed the largest Lyapunov exponent and a coarse-grained Shannon entropy for ensembles of trajectories in the H\'enon-Heiles potential and for test-particle orbits in live $N$-body realisations of a Plummer model. We compared the dependence of both quantities on the orbital energy as well as on the particle number for the $N$-body case in particular.}
   {
   In the H\'enon-Heiles system, the Shannon entropy follows the transition from weak to widespread chaos and exhibits an energy dependence that closely mirrors that of the largest Lyapunov exponent. For test-particle orbits in live $N$-body potentials, both diagnostics indicate stronger chaos for more tightly bound trajectories. However, their dependence on $N$ differs: the largest Lyapunov exponent remains nearly constant over the explored range of particle numbers, whereas the Shannon entropy decreases monotonically as $N$ increases.}
   {
   These results show that the Shannon information entropy could serve as a complement to the largest Lyapunov exponent and could also better capture changes in global phase-space mixing, especially in systems where the leading Lyapunov exponent alone is not sufficiently informative. It provides a promising alternative for diagnosing chaos when tangent-space dynamics is unavailable, prohibitively expensive, or slowly convergent. In particular, because this entropy measure relies only on time series of phase-space coordinates, it is naturally suited to systems with densely sampled trajectories, such as minor bodies in the Solar System.
   }

   \maketitle
\section{Introduction}\label{intro}
Gravitational systems, even for relatively low numbers of degree of freedom, are known to be characterised by the coexistence of regular and chaotic orbits, the latter typically defined as exhibiting strong sensitivity to initial conditions \citep{2002ocda.book.....C}. The main open issues surrounding chaos in the gravitational $N$-body problem can be summarised as follows: (1) the relation between Hamiltonian chaos and the onset of dynamical instabilities in the few body systems \citep{1992Natur.357..569M,2008ApJ...683.1207B,2023PhRvX..13b1018M}; (2) the dependence of the extent of chaos on the specific orbital structure of a given energy landscape in galactic potentials (\citealt{1998MNRAS.298....1C,2001A&A...367..443K,2003MNRAS.345..727K}); and (3) the scaling of chaos with the number of degrees of freedom of the systems at hand (\citealt{1971JCoPh...8..449M,1986A&A...160..203G,1991ApJ...374..255K,1993ApJ...415..715G,2001PhRvE..64e6209K,2019MNRAS.484.1456E,2019MNRAS.489.5876D,2023MNRAS.526.5791P}).\\
\indent Usually, the degree of chaos in an autonomous dynamical system that can be expressed in Hamiltonian from $\mathcal{H}({x}_1,p_1,...,{x}_N,p_N)={\rm const}$ is estimated by evaluating numerically its largest Lyapunov exponent  (e.g. see \citealt{1992rcd..book.....L}), formally defined by
 \begin{equation}\label{eq:lmaxformal}
\lambda_{\rm max} =\lim_{t\to\infty} \lim_{||\mathbf{W}_0||\to0}\frac{1}{t}\ln\frac{||\mathbf{W}(t)||}{||\mathbf{W}_0||}~.
\end{equation}
In the equation above, $\mathbf{W}=({\delta x}_1,\delta p_1,...,{\delta x}_N,\delta p_N)$ denotes the state vector in the tangent space, $\mathcal{S}_T$, to the phase space, $\mathcal{S}$, of the system. The dynamics of the tangent vectors is given by 
\begin{equation}\label{eq:tangent2}
{\delta\dot{\mathbf{x}}}=\delta\mathbf{p};\quad {\delta\dot{\mathbf{p}}}=-\mathbf{D}^2_V(\mathbf{x})\delta\mathbf{x},
\end{equation}
where $\mathbf{x}=(x_1,...,x_N)$, with analogous definitions for $\delta\mathbf{x}$, $\mathbf{p}$, and $\delta\mathbf{p}$. We also have
\begin{equation}\label{eq:matrix}
\mathbf{D}^2_V(\mathbf{x})_{jk}=\left.\frac{\partial^2 V(\mathbf{x})}{\partial x_j\partial x_k}\right |_\mathbf{x};\quad j,k=1,2,...,N,
\end{equation}
where $V(\mathbf{x})$ is the potential part of the Hamiltonian $\mathcal{H}=\mathbf{p}^2/2+V(\mathbf{x})$. 

\REV{If computing the Hessian matrix $D^2 V$ is particularly cumbersome or if the system is non-autonomous (i.e. $\mathcal{H}$ depends explicitly on time), we would typically replace the norm of the tangent vector, $\|\mathbf{W}(t)\|$, in Eq.~(\ref{eq:lmaxformal}) with the phase-space distance between two initially nearby trajectories, $\xi(t \mid x_0)$ and $\xi'(t \mid x_0 + \delta x_0)$ \citep{benettin1980}. In this approach, the numerically estimated $\lambda_{\rm max}$ may depend on the initial normalisation of $\mathbf{W}_0$ unless periodic renormalisation is applied.

This procedure consists of evolving the two trajectories for a finite time interval and then redefining the perturbed trajectory so that its separation from the reference trajectory is reset to the initial magnitude, $\delta x_0$, while preserving its direction. Explicitly, at each renormalisation step, we have
\begin{equation}
\xi'(t) \;\rightarrow\; \xi(t) + \delta x_0 \,\frac{\xi'(t) - \xi(t)}{\|\xi'(t) - \xi(t)\|}.
\end{equation}

This renormalisation ensures that the evolution remains within the linear (tangent) regime; namely, that the perturbation stays infinitesimal. Without such a procedure, the separation between trajectories eventually saturates at a characteristic scale of the system (e.g. the system size or the typical impact parameter between particles), thereby invalidating the measurement of exponential divergence, as noted by several authors \citep{1993ApJ...415..715G,valluri1999,2019MNRAS.484.1456E,2023MNRAS.526.5791P}.}

More generally, the full Lyapunov spectrum can be computed by propagating the variational equations up to a time, $\tau$, after which the tangent vectors are orthonormalised (e.g. via a Gram–Schmidt procedure) to prevent their alignment with the most unstable direction \citep{1993PhRvE..47.3686W,1997Nonli..10.1063C,2011A&A...533A...2Q}. The finite-time Lyapunov exponents, $\lambda_i(\tau)$, are then obtained by accumulating the logarithmic growth factors associated with the orthonormalisation via
\begin{equation}\label{eq:spectrum1}
\lambda_i(\tau)=\frac{1}{\tau}\left(\lambda_{i-1}(\tau)+\ln\Bigg|\Bigg|\mathbf{z}_i-\sum_{j=1}^i\langle\mathbf{z}_i,\mathbf{z}_{j-1}^\prime\rangle\mathbf{z}_{j-1}^\prime\Bigg|\Bigg|\right),
\end{equation}
up to the desired precision dictated by the chosen convergence criterion. In Eq.~(\ref{eq:spectrum1}), $\mathbf{z}_i=(\delta x_i,\delta p_i)$ denotes the $i$-th deviation (tangent) vector, initially normalised to unity, and $\lambda_0=0$. 

\REV{
The largest Lyapunov exponent is only the leading member of the Lyapunov spectrum. While $\lambda_{\max}$ controls the fastest infinitesimal separation between nearby trajectories, the subleading positive exponents quantify additional unstable directions in tangent space \citep{ISHIDA20065035,ZHOU2020109981}. Equivalently, the sum of the first $k$ exponents gives the exponential growth rate of an infinitesimal $k$-dimensional volume element. Therefore, two systems with comparable $\lambda_{\max}$ may still differ substantially in the dimensionality and isotropy of their instability. This distinction is particularly relevant for Hamiltonian systems with mixed phase space, where local stretching, transport across phase space, and finite-time trapping near regular structures need not be controlled by a single exponent.}

Using the largest Lyapunov exponent as a metric of chaos to investigate its scaling in the continuum limit of the $N-$body problem (i.e. $N\to\infty$ with asymptotically vanishing individual particle mass), leads to puzzling findings. For a given model, determined by a potential density pair supported by an equilibrium phase-space distribution, $f$, single particle trajectories, largely independent of their specific energies, will increasingly resemble their parent realisations in the smooth potential of the continuum model, while their largest Lyapunov exponents depend weakly on the number of simulation particles, $N$. 

\cite{2001PhRvE..64e6209K} carried out frozen $N$-body experiments (i.e. test particles propagated through the potential generated by a distribution of $N$ fixed particles sampling a given density profile) and concluded that chaos associated with discreteness effects in the $N$-body problem should be viewed very differently from the chaos associated with a bulk, possibly non-integrable, potential. More recently, \cite{2019MNRAS.489.5876D} studying test orbits in live direct $N$-body simulations with $N$ up to ${\approx} 10^5$ reached similar conclusions. In particular (see their Figs. 7 and 12) the weak scaling of orbital chaoticity with the number of degrees of freedom (i.e. number of particles at a fixed normalisation) has a different trend for different energies. 

With respect to the largest Lyapunov exponent, $\lambda_{\rm max}$, of the full $N$-body problem, whereas some studies suggest that it increases for increasing $N$  \citep[see][]{1993ApJ...415..715G,2002ApJ...580..606H,2022A&A...659A..86P,2023AIPC.2872e0003P,asano2026}, some others (\citealt{2019ApJ...870..128B,2019MNRAS.489.5876D,2020MNRAS.494.1027D}) indicate instead that $\lambda_{\rm max}$ is inversely proportional to $N$, at least for $N\gtrsim 10^4$. This was previously suggested in the semi-analytical estimations of \cite{1986A&A...160..203G} and \cite{2009A&A...505..625G}. These manifestly contradicting findings are not reconciled at present. The use of a regularised version of the Coulombian $1/r$ potential, such as the usual $1/\sqrt{r^2+\epsilon^2}$, can only partially explain why models with softened interactions have systematically lower Lyapunov exponents due to the absence of hard scattering events\footnote{We note that, \cite{1986A&A...160..203G} when evaluating their $\lambda_{\rm max}\propto N^{-1/3}$ relation imposed a cut-off on the tail of the stochastic force distribution acting on each test mass, de facto corresponding to softening the gravitational interaction below a fixed scale length (D.Heggie, private communication).} among particles. Moreover, convergence studies at fixed system parameters with decreasing values of the softening length $\epsilon$, yield similar values of the Lyapunov exponents below some critical $\epsilon$ (see Fig.~2 of \citealt{2001PhRvE..64e6209K} and Fig.~6 of \citealt{2019MNRAS.489.5876D}). 

These apparently conflicting results indicate that the use of the Lyapunov exponent alone might not be enough to characterise chaos in gravitational $N$-body systems. Of course, up to this point we have discussed the largest $\lambda$. The full spectrum may differ substantially between realisations with different $N$ and similar $\lambda_{\rm max}$. For example \cite{2025A&A...693A..53D} showed that in the three-body problem with relativistic corrections included in the force calculations, the largest Lyapunov exponents can be comparable to (or even smaller than) their classical counterparts, while the full spectrum may differ significantly between the two cases. 

It is therefore worth exploring other metrics of chaos in addition to the Lyapunov exponents and related quantities such as the small alignment indexes (SALI, \citealt{2001JPhA...3410029S}), generalised alignment indexes (GALI, \citealt{2007PhyD..231...30S}), or the  mean exponential growth factor
of nearby orbits (MEGNO, \citealt{2000A&AS..147..205C,2011MNRAS.414L.100M}). For example, several studies in the context of simple dynamical systems (e.g. see \citealt{1997JSP....88....1C,PhysRevE.61.1337,SHIOZAWA2024129531}) used the Kolmogorov-Sinai \citep[KS,][]{kolmogorov0,kolmogorov,sinai} entropy $\mathcal{S}_{\rm KS}$ \citep[see][]{1992rcd..book.....L}. Once a coarse graining of time and phase-space is defined, the KS entropy quantifies the rate of information production along a trajectory, as inferred from the conditional probabilities that the trajectory occupies a given phase-space cell, $j$, at time, $t_{n+1}$, given its history up to time $t_n$. \REV{Remarkably, \cite{1977RuMaS..32...55P} showed that for smooth dynamical systems with an invariant measure, such as Hamiltonian systems obeying Liouville’s theorem, the KS entropy is bounded from above by the sum of the positive Lyapunov exponents,
\begin{equation}\label{kse}
\mathcal{S}_{\rm KS} \leq \mathcal{S}^+_{\rm KS} = \sum_{\lambda_i>0}\lambda_i.
\end{equation}
When the dynamics is sufficiently chaotic, this bound becomes an equality, known as Pesin’s identity. In systems with mixed phase space, where regular and chaotic regions coexist, this relation should be understood as applying to the chaotic component in the asymptotic limit.}

Another commonly used entropy measure is the Shannon information entropy \citep{1949mtc..book.....S}. Given a partition of phase space into cells $k$, with occupation probabilities, $p_k(t)$, for instance, estimated from an ensemble of trajectories at a fixed time $t$), the Shannon entropy is
\begin{equation}\label{eq:shannon}
        \mathcal{S}_{\rm Sh}(t) = -\sum_k p_k(t)\,\ln p_k(t),
\end{equation}
which quantifies how broadly the ensemble is distributed over the chosen coarse-grained representation. The Shannon and KS entropy are conceptually similar in that both depend on coarse graining and capture information content. However, $\mathcal{S}_{\rm Sh}(t)$ is an instantaneous property of a probability distribution (and can be tracked in time if the distribution evolves), whereas $\mathcal{S}_{\rm KS}$ is an asymptotic entropy rate associated with the time evolution of a single trajectory, defined from the statistics of sequences of its visits in phase space cells.

\REV{The use of entropy-based quantities as indicators of relaxation and chaos has been rather successful in dynamical astronomy \citep[e.g.][]{romeo1990}}. In a series of studies \citep{nunez1996,cincotta1999,giordano2018,cincotta2021}, the information entropy was employed to quantify the dynamical stability of ensembles of trajectories in the restricted three-body problem and to identify periodic and chaotic patterns in astronomical data.

These diagnostics are complementary to frequency map analysis, which characterises orbital regularity through the Fourier decomposition of time series and the (in)stability of the associated fundamental frequencies. \REV{Originally introduced in planetary dynamics \citep{laskar1990}, frequency map analysis has since been widely adopted in accelerator beam dynamics \citep{shatilov2011} and galactic dynamics \citep{valluri2012,vasiliev2013,bajkova2023,woudenberg2025}. More recently, \citet{hyman2025} introduced a framework for orbital-complexity analysis based on permutation entropy and statistical complexity measures, while \citet{canbaz2025} proposed a novel indicator called a permutation entropy of the power spectrum, combining permutation entropy with Fourier-based methods. Similarly, \citet{bajkova2025} computed the information entropy of the discrete Fourier transform of time series of radial Galactic distances to classify the chaoticity of globular cluster orbits perturbed by the Galactic bar.}

In this work, we investigate the relationship between different chaos indicators by comparing the largest Lyapunov exponent and the Shannon entropy in two idealised systems: a test particle in the H\'enon-Heiles potential and in the potential generated by $N$ particles distributed according to a Plummer profile. We assess the degree to which these indicators yield consistent classifications of regular and chaotic dynamics, and we identify regimes where their interpretations diverge. We also examine the dependence of these indicators on the particle number, $N$, with the aim of clarifying the apparently contradictory results on the relation between chaoticity and the number of degrees of freedom in the $N$-body problem.

The rest of the paper is structured as follows. In Sect. \ref{sec:methods}, we introduce the governing equations, discuss the numerical integration and describe the procedure to evaluate the Lyapunov exponents and the information entropy. In Sect. \ref{sec:results}, we present our numerical simulations and discuss the results. Finally, in Sect. \ref{sec:conclusions} we draw our conclusions and interpret our findings in light of previous work.

\section{Methods}\label{sec:methods}
\subsection{Models}
In this work, we consider two models, specifically, the \citet[hereafter HH]{henon1964} system and a spherical $N$-body system in which the motion of a single particle is studied within the time-dependent gravitational potential generated by the remaining $N-1$ particles. 

\subsubsection{H\'enon-Heiles system}
The HH system is defined by the Hamiltonian,
\begin{equation}\label{eq:HH}
\mathcal{H}_{\rm HH}=\frac{p_x^2+p_y^2}{2}+\frac{x^2+y^2}{2}+x^2y-\frac{y^3}{3},
\end{equation}
and it was originally conceived as a toy model for the dynamics in the meridional plane of an axisymmetric galactic potential. It has since become a paradigmatic example in nonlinear dynamics, exhibiting a mixed phase space; namely, a nontrivial coexistence of regular and chaotic orbits on the same energy surface \citep{weiss2003}. In this work, we adopted the conventional rescaling in which the particle mass $m$ and the coefficients of all terms in the 2D potential are set to unity. With this normalisation, the critical energy above which all trajectories escape is $E_{\rm crit} = 1/6$.
\begin{figure}
    \centering
    \includegraphics[width=\columnwidth]{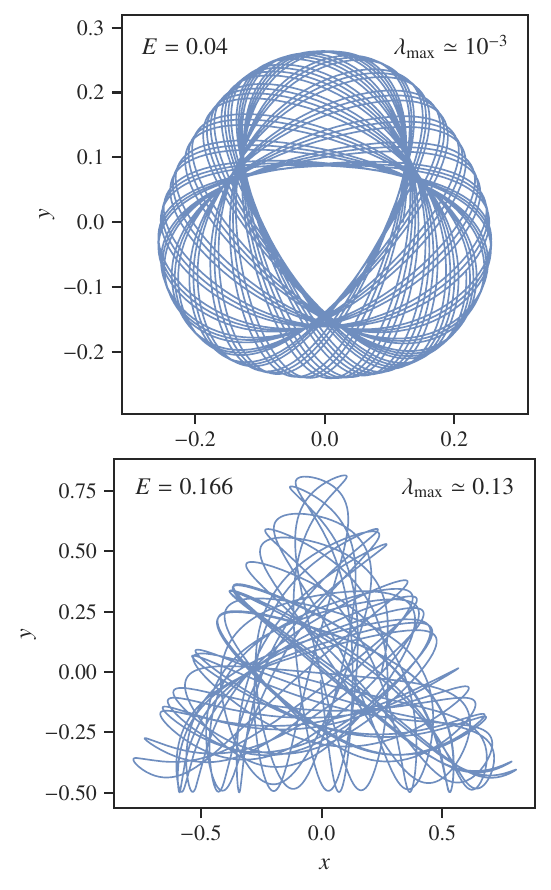}
    \caption{Illustration of two trajectories in the HH system. Top panel: Low-energy, regular trajectory with a nearly zero largest Lyapunov exponent, $\lambda_\mathrm{max}$. Bottom panel: High-energy, chaotic trajectory with finite $\lambda_\mathrm{max}$.
    }
    \label{fig:hhexample} 
\end{figure}
The equations of motion derived by Eq. (\ref{eq:HH}) and the associated variational equations, defined by Eq.~(\ref{eq:matrix}), is expressed as
\begin{align}\label{eq:eomhh}
& \ddot{x} = -x-2xy ,\\
& \ddot{y} = -x^2+y^2-y,
\end{align}
and
\begin{align}
&  \ddot{\delta x} = -(1+2y) \,\delta x-2x\,\delta y, \\
&  \ddot{\delta y} = -2x\,\delta x -(1-2y)\,\delta y,
\end{align}
respectively. Figure~\ref{fig:hhexample} illustrates a typical regular and chaotic orbit in the HH system, corresponding to $E=0.04$ and $E=0.16$.
\subsubsection{$N$-body system}
We probe the dynamics of individual orbits in spherical equilibrium self gravitating $N$-body models evolving  test particles in the time-dependent gravitational potential exerted by the remaining $N-1$ bodies. The equation of motion for the $i$-th particle of mass, $m$, is
\begin{equation}\label{eq:eom}
\ddot{{\mathbf r}}_i=-Gm\sum_{i\neq j=1}^N\frac{{\mathbf r}_i-{\mathbf r}_j}{r_{ij}^3},
\end{equation}
where $G$ is the gravitational constant and $r_{ij}=||\mathbf r_i-\mathbf r_j||$. The associated variational equations in tangent space, required for the computation of Lyapunov exponents (see, e.g. \citealt{2016MNRAS.459.2275R}), is expressed as
\begin{equation}\label{eq:var}
        \ddot{\delta\mathbf r}_i=-Gm\sum_{\substack{j=1 \\ j\neq i}}^N\left[\frac{\delta\mathbf r_i-\delta\mathbf r_j}{r_{ij}^3}
        -3(\mathbf r_i-\mathbf r_j)\frac{\langle\delta\mathbf r_i-\delta\mathbf r_j,\mathbf r_i-\mathbf r_j\rangle}{r_{ij}^5}\right].
\end{equation}
In the $N$-body experiments, the initial positions and velocities of the $N$ particles are drawn from the widely adopted \citet{plummer1911} model, with total mass, $M$, and scale radius, $a$, and defined by the following spherical density-potential pair:
\begin{equation}\label{eq:plummerrho}
    \rho(r)=\frac{3 M}{4\pi a^3}\left( 1 + \frac{r^2}{a^2} \right)^{-5/2};\quad \Phi(r)=-\frac{GM}{\sqrt{r^2+a^2}},
\end{equation}
supported by the ergodic phase-space distribution function,
\begin{equation}
f(E)=\frac{\sqrt{2}}{378\pi^2Ga^2\sigma_0}\left(\frac{-E}{\sigma_0^2}\right)^{7/2}.
\end{equation}
Here, $E$ denotes the specific energy of a particle and $\sigma_0=\sqrt{GM/6a}$ is the central velocity dispersion. All quantities are expressed in $N$-body units, such that $G=M=a=1$ and each particle has equal mass, $m=M/N$. Consequently, the typical crossing time $t_\mathrm{c}\equiv\sqrt{a^3/GM}=1$. We simulate different sets of initial conditions with $N$ between $10^4$ and  $10^6$. As $N$ increases, the total mass is held fixed, consequently, the individual particle mass decreases accordingly.

We approximate the dynamics by evolving $N-1$ independent ``field'' particles in the smooth potential (\ref{eq:plummerrho}), while computing the trajectory and associated tangent dynamics of a ``test'' particle according to Eqs.~(\ref{eq:eom}–\ref{eq:var}). In this framework, the variational vectors of the field particles, $\ddot{\delta\mathbf r}_j$, evolve in the Plummer potential according to 
\begin{equation}\label{eq:varplummer}
\ddot{\delta\mathbf r}_j=-Gm\left[\frac{\delta\mathbf r_j}{(r_{j}^2+a^2)^{3/2}}
        -3\mathbf r_j\frac{\langle\delta\mathbf r_j,\mathbf r_j\rangle}{(r_{j}^2+a^2)^{5/2}}\right].
\end{equation}

\subsection{Numerical integration}
For both the HH system and the $N$-body system, we integrated the equations of motion, Eqs.~(\ref{eq:eomhh}) and~(\ref{eq:eom}), together with their associated variational equations \citep{skokos2010}, using the fourth-order symplectic integration scheme discussed in \citet[see also \citealt{yoshida1990,yoshida1993,kin91} and references therein]{laskar2001}. In all runs, we adopted a fixed time step, $\Delta, t$ that depends on the initial orbital energy. The time step was determined through preliminary trial-and-error integrations, requiring that over an integration time of $5\times10^3$ time units the relative energy error remains below $10^{-9}$ in double precision for the HH system. In the $N$-body runs, for the dimensionless equations of motion considered here, the resulting optimal timesteps typically lie in the range $2\times10^{-5}\leq\Delta t\leq10^{-2}$. As we are following individual test trajectories in time dependent $N$-body potentials, we did not employ any softening or regularisation of the gravitational interactions.

\subsection{Evaluation of the Lyapunov exponents}
Evaluating the largest Lyapunov exponent of a given system by simply time-dependently computing (\ref{eq:lmaxformal}) up to a desired convergence is typically unfeasible as it is in principle plagued by round-off errors resulting in a rapid blow-up of the numerical solution.  Following \cite{1976PhRvA..14.2338B} in the numerical calculations we obtain $\lambda_{\rm max}$  as the limit of 
 \begin{equation}\label{lmax}
\lambda_{\rm max}(t) =\frac{1}{L\Delta t}\sum_{k=1}^L\ln\frac{||\mathbf{W}(k\Delta t)||}{||\mathbf{W}_0||}~,
\end{equation}
for a (large) time $t=L\Delta t$ with the additional precaution of periodically re-normalising $\mathbf{W}$ to its initial magnitude (see e.g. \citealt{2010LNP...790...63S} and references therein). When computing $\lambda_{\rm max}$ for an individual particle in $N-$body experiments, we also average over different experiments where the other $N-1$ background particles are sampled from the same model in a similar fashion to what done in \cite{2025A&A...698A..28S} for particles in time-dependent gravitational potential subjected to noise.
\begin{figure}
    \centering
    \includegraphics[width=1\columnwidth]{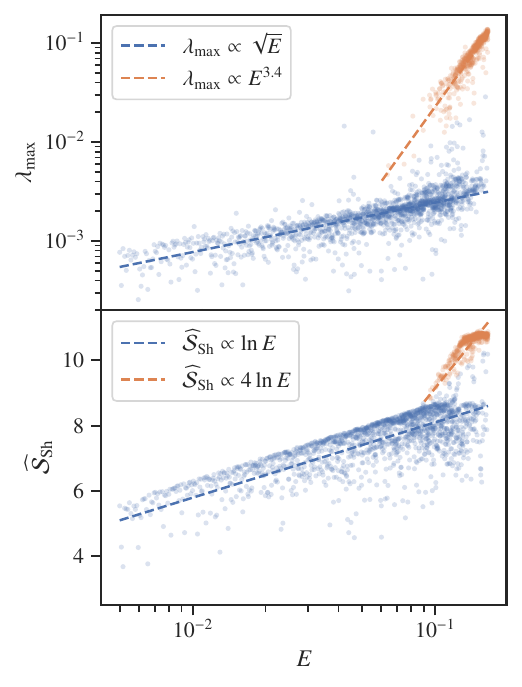}
    \caption{Largest Lyapunov exponent, $\lambda_\mathrm{max}$ (top panel), and Shannon entropy, $\widehat{\mathcal{S}}_{\rm Sh}$ (bottom panel), versus energy for 2000 realisations of the HH system. In the top panel, the blue and red dashed lines mark the trends $\lambda_{\rm max}\propto \sqrt{E}$ and $\lambda_{\rm max}\propto E^{3.4}$ reported by \cite{2003JETPL..77..642S}. In the bottom panel, the curves $\widehat{\mathcal{S}}_{\rm Sh} \propto \ln E$ and $\widehat{\mathcal{S}}_{\rm Sh} \propto 4\ln E$ are shown for comparison. Both $\lambda_\mathrm{max}$ and $\widehat{\mathcal{S}}_{\rm Sh}$ increase with energy, with a change in slope around $E\simeq 0.08$, marking the transition from predominantly regular to chaotic trajectories.
    }
    \label{fig:lmaxHH} 
\end{figure}
\begin{figure}
        \centering
        \includegraphics[width=1\columnwidth]{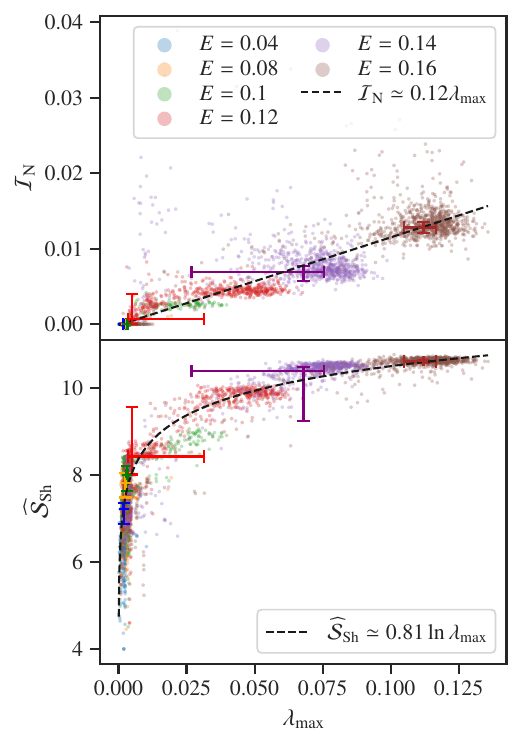}
        \caption{Average mutual information entropy, $\mathcal I_{\rm N}$ (top panel), and Shannon entropy (bottom panel) versus the largest Lyapunov exponent, $\lambda_{\rm max}$, for different values of the energy, $E$, in the range $0.04\leq E \leq 0.16$ in the HH system. Error bars indicate the $1\sigma$ scatter over the ensemble, shown by the coloured points. In both panels, dashed lines represent the best-fitting relation between the entropy and $\lambda_{\rm max}$.}
        \label{fig:scatter2}
\end{figure}
\begin{figure*}
    \centering
    \includegraphics[width=\textwidth]{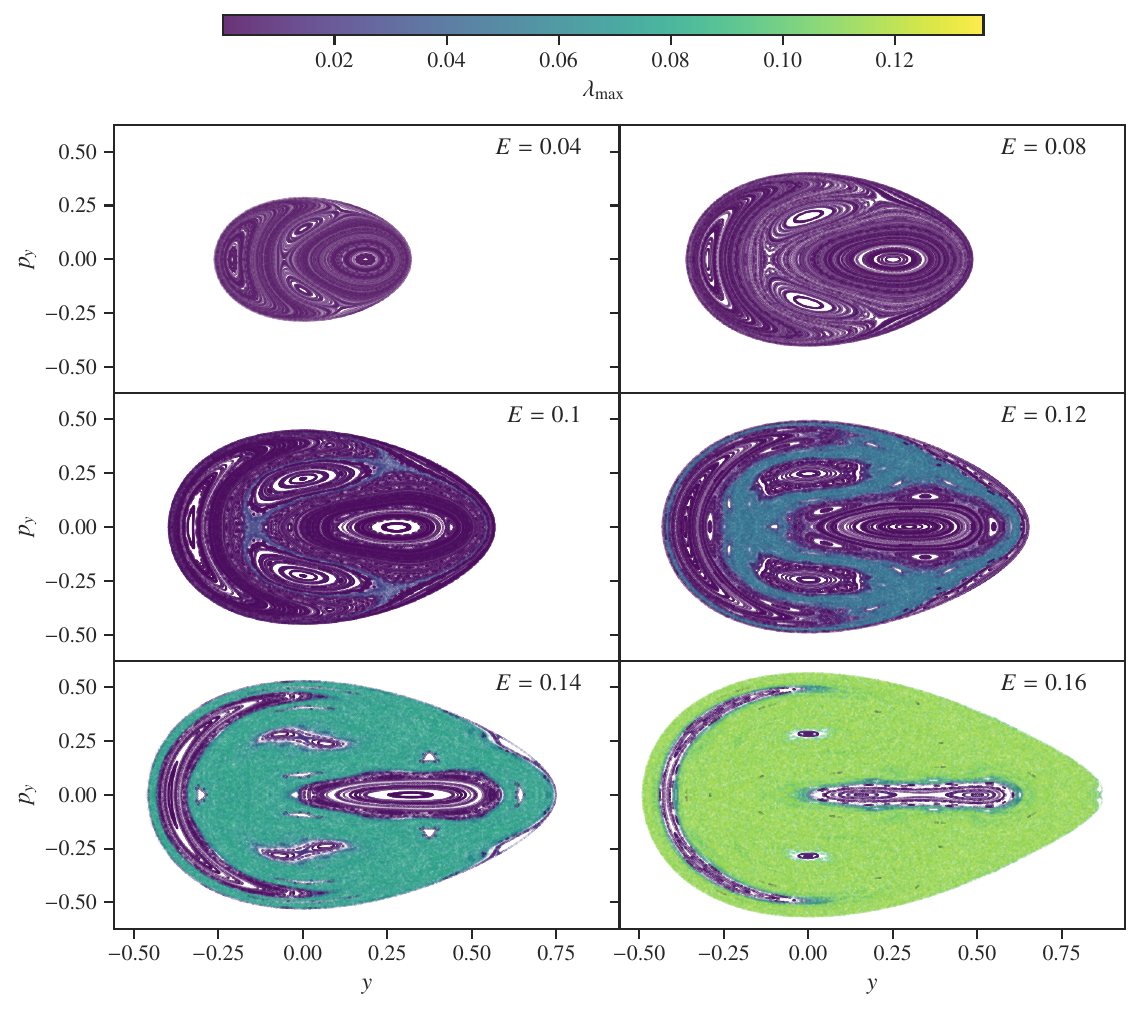}
    \caption{Poincar\'e sections $(y, p_y)$ of the HH Hamiltonian for different values of the energy, $E$, defined by crossings of the plane $x=0$. The colour coding represents the largest Lyapunov exponent, $\lambda_{\rm max}$, of the orbits.
    }
    \label{fig_poincare_L} 
\end{figure*}
\begin{figure*}
    \centering
    \includegraphics[width=\textwidth]{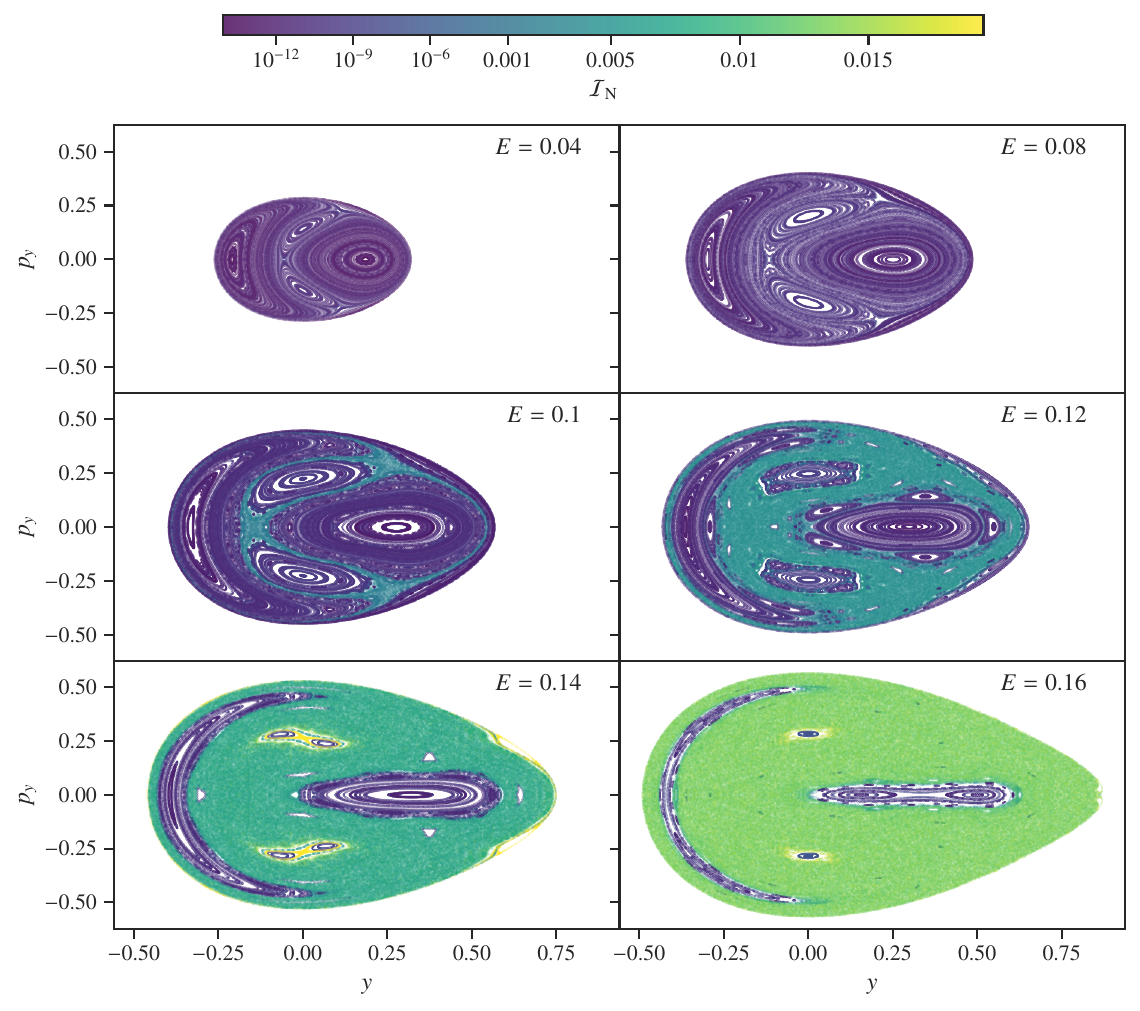}
    \caption{Poincar\'e sections $(y, p_y)$ of the HH Hamiltonian for different values of the energy, $E$, defined by crossings of the plane $x=0$. The colour coding represents the mutual information entropy, $\mathcal{I}_{\rm N}$ (Eq.~\ref{eq:entropynunez}), of the orbits.
    }
    \label{fig_poincare_H} 
\end{figure*}
\subsection{Information entropy}
We computed the information entropy in two distinct ways. First, for individual orbits we evaluate Eq.~(\ref{eq:shannon}) from the empirical phase-space density in the full state vector $(x,y,p_x,p_y)(t)$, obtained by coarse-graining the trajectory samples into a multidimensional histogram. Using $\hat p_k$ to denote the fraction of stored samples that fall in bin, $k$, our estimator is
\begin{equation}\label{entropy_sum}
        \widehat{\mathcal{S}}_{\rm Sh} \equiv -\sum_k \hat p_k \ln \hat p_k,
\end{equation}
where the sum runs over all bins (with the convention that empty bins contribute zero).
To mitigate the dependence on an arbitrary bin choice, we estimated a reference bin width in each coordinate using the Freedman-Diaconis rule and then scanned a range of nearby binnings to verify the stability of the inferred entropy. We enforced a minimum sampling quality by requiring an average occupancy of at least five samples per non-empty bin (i.e.\ $N_{\rm samp}/N_{\rm occ}\ge 5$), where $N_{\rm samp}$ is the number of stored phase-space samples and $N_{\rm occ}$ is the number of occupied bins. For consistent entropy comparisons, a common binning is adopted for all trajectories within a given comparison set. We also explored nearby binning configurations and verified that the qualitative trends, as well as the quantitative values within uncertainties, remain unchanged. In this definition, $\widehat{\mathcal{S}}_{\rm Sh}$ quantifies how broadly the trajectory populates the accessible phase space (at the chosen coarse-graining). This is therefore a measure of phase-space exploration, rather than of instantaneous information-production rate.
\begin{figure*}
        \centering
        \includegraphics[width=1\textwidth]{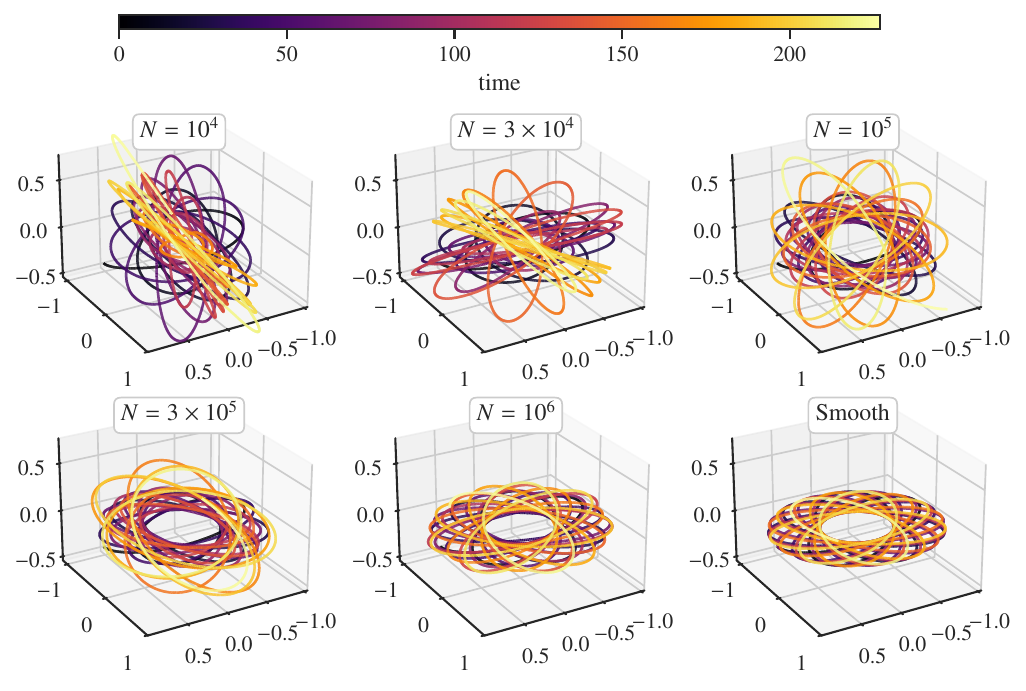}
        \caption{Tracer particle orbit with initial energy $E\approx -0.66$ in different $N$-body realisations of an isotropic Plummer model with increasing $N$, and in the smooth-potential (continuum) limit. The colour coding represents time in units of $t_\mathrm{c}$. As $N$ increases, the orbit approaches its continuum counterpart.}
        \label{fig:orbitnbody}
\end{figure*}

Second, we evaluated the mutual information entropy introduced by \citet[][see also \citealt{fraser1986}]{nunez1996}, which we denote with $\mathcal{I}_{\rm N}$. Although its formal definition resembles that of a mutual information Shannon entropy, the quantity $\mathcal{I}_{\rm N}$ is not a mutual information measure in the probabilistic sense.
Instead, it is constructed by calculating the Shannon entropy of normalised arc-length weights along two trajectories in phase space and by combining the resulting entropies in a way that is sensitive to the dynamical de-correlation between nearby realisations.
Given a reference trajectory $\xi(t)$ and a nearby realisation, $\xi'(t)$, initialised at a separation of $\|\xi'(0)-\xi(0)\|=d_0$, \citet{nunez1996} define
\begin{equation}\label{eq:entropynunez}
        \mathcal{I}_{\rm N}(t)= \mathcal{S}_{\rm N}(\xi,\xi';t) - \frac{1}{2}\bigg[\mathcal{S}_{\rm N}(\xi;t) + \mathcal{S}_{\rm N}(\xi';t)\bigg],
\end{equation}
where the entropies are computed from normalised path-length rates along the curves.
For a single trajectory, $\xi(t)$, we define the instantaneous speed in phase space,
\begin{equation}
        \dot \ell_\xi(t) = \|\dot\xi(t)\|,
\end{equation}
the accumulated path length up to time, $t$,
\begin{equation}
        L_\xi(t) = \int_0^t \dot\ell_\xi(t')\,{\rm d}t',
\end{equation}
and the corresponding normalised length density,
\begin{equation}
        \varepsilon_\xi(t) = \frac{\dot\ell_\xi(t)}{L_\xi(t)}
.\end{equation}
The path-length entropy for an individual trajectory is then the Shannon entropy of this time-density,
\begin{equation}\label{eq:snunez}
        \mathcal{S}_{\rm N}(\xi;t) = -\int_0^t \varepsilon_\xi(t') \ln \varepsilon_\xi(t')\, {\rm d}t'.
\end{equation}
For the joint entropy estimator $\mathcal{S}_{\rm N}(\xi,\xi';t)$, we define
\begin{align}
&       \dot\ell_{\xi\xi'}(t) = \sqrt{\dot\ell_\xi^2(t)+\dot\ell_{\xi'}^2(t)},\\
&       L_{\xi\xi'}(t) = \int_0^t \dot\ell_{\xi\xi'}(t')\,{\rm d}t',\\
&       \varepsilon_{\xi\xi'}(t) = \frac{\dot\ell_{\xi\xi'}(t)}{L_{\xi\xi'}(t)},
\end{align}
and substitutes $\varepsilon_{\xi\xi'}$ into Eq.~(\ref{eq:snunez}) to obtain $\mathcal{S}_{\rm N}(\xi,\xi';t)$.
In contrast to $\widehat{\mathcal{S}}_{\rm Sh}$, $\mathcal{I}_{\rm N}(t)$ is sensitive to dynamical divergence: as the perturbed trajectory deviates from the reference one, the joint and marginal length densities evolve differently, producing a characteristic growth and eventual saturation of $\mathcal{I}_{\rm N}(t)$ in chaotic regimes \citep{nunez1996}. In this sense, $\mathcal{I}_{\rm N}$ plays a role analogous to that of a Lyapunov exponent as a diagnostic of instability, while remaining a finite-time quantity. A practical distinction is that the mutual information entropy as defined by \citet{nunez1996} cannot, in general, be computed from tangent-vector dynamics alone: its definition depends on the accumulated path length of a genuinely distinct trajectory, which is not preserved under the renormalisation steps required by the Lyapunov-exponent computation algorithm.
\begin{figure}
        \centering
        \includegraphics[width=1\columnwidth]{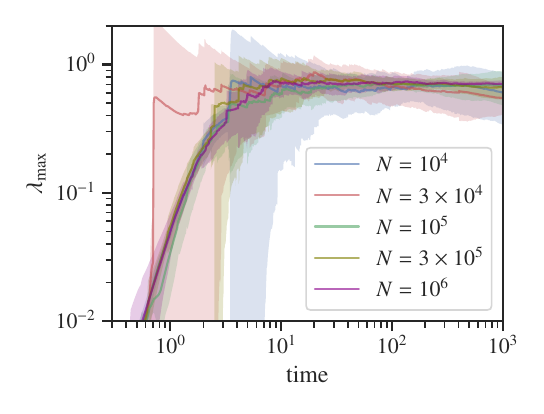}
        \caption{Evolution of the largest Lyapunov exponent $\lambda_{\rm max}$ for different values of $N$ in the $N$-body system. Solid lines show the mean over 10 realisations and the shaded regions indicate the $1\sigma$ dispersion.}
        \label{fig:lambdaN}
\end{figure}
\begin{figure}
        \centering
        \includegraphics[width=1\columnwidth]{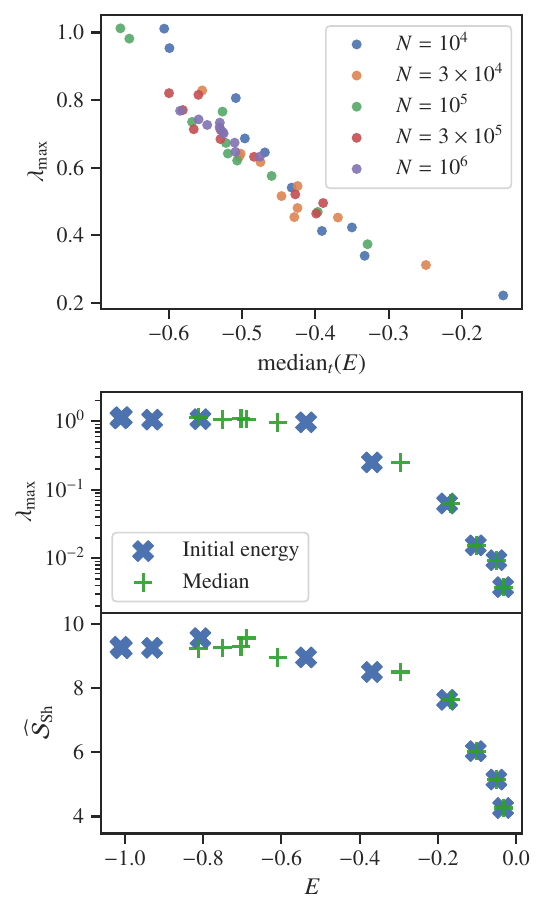}
        \caption{Top panel: Largest Lyapunov exponent, $\lambda_\mathrm{max}$, versus the orbital energy, $E$, of the test particle in the $N$-body system, where $E$ is taken as the median along the trajectory. Colors indicate the number of particles, $N$. Lower panels: $\lambda_\mathrm{max}$ (upper sub-panel) and Shannon entropy, $\widehat{\mathcal{S}}_{\rm Sh}$ (lower sub-panel) ,versus the initial (blue crosses) and median (green crosses) orbital energy in a realisation of an isotropic Plummer model at fixed $N=10^5$. Both $\lambda_\mathrm{max}$ and $\widehat{\mathcal{S}}_{\rm Sh}$ decrease with increasing orbital energy and approach a saturation level for the most tightly bound orbits.}
        \label{fig:lambdae}
\end{figure}
\section{Numerical experiments and results}\label{sec:results}
\subsection{HH system}
It is well known that in the HH potential the largest Lyapunov exponent $\lambda_{\rm max}$, averaged over an energy surface, increases with $E$. Early computations by \cite{1976PhRvA..14.2338B} suggested an $\exp{E}$ trend for both $\lambda_{\rm max}$ and the KS entropy, $S_{\rm KS}$. Subsequent studies, for example \cite{2003JETPL..77..642S} reported a broken power-law trend instead, with a sharp change in slope at $E \approx 0.08$ for the averaged Lyapunov exponent.

In the energy range $10^{-3}\leq E<1/6$, we evaluated both the largest Lyapunov exponent and the Shannon entropy on a grid of $10^3$ realisations. The top panel of Fig.~\ref{fig:lmaxHH} shows $\lambda_{\rm max}$ as a function of orbital energy $E$. Orange and blue points highlight the two regimes, indicated by the corresponding dashed lines, where $\lambda \propto \sqrt{E}$ and $\lambda \propto E^{3.4}$, respectively. The Shannon entropy, $\widehat{\mathcal{S}}{\rm Sh}$, for the same orbits is shown in the bottom panel, and likewise exhibits a broken trend with $E$, consistent with $\widehat{\mathcal{S}}_{\rm Sh} \propto \ln E$ for $E \lesssim 0.08$ and $\widehat{\mathcal{S}}_{\rm Sh} \propto \ln E^4$ for $E \gtrsim 0.08$. We note that, in principle, for a given value of $E$, due to the coexistence of wildly chaotic and nearly regular orbits, the values of the Lyapunov exponents for different trajectories can differ significantly, resulting in highly non-trivial distributions $P(\lambda)$ (e.g. \citealt{PhysRevE.78.066204}).

In Fig.~\ref{fig:scatter2}, we present the mutual information entropy, $\mathcal{I}_N$ (top panel), in addition to the Shannon entropy $\widehat{\mathcal{S}}_{\rm Sh}$ (bottom panel) against the mean Lyapunov exponent. We find that $\mathcal{I}_N$ scales linearly with $\lambda_{\rm max}$, while the Shannon entropy is compatible with a $\widehat{\mathcal{S}}_{\rm Sh}\propto\ln\lambda_{\rm max}$ relation, in agreement with the picture emerging from Fig.~\ref{fig:lmaxHH}. 

The correspondence between the mutual information entropy and the Lyapunov exponent can be visualised by colour-coding these indicators in the Poincar\'e sections of the trajectories at fixed values of $E$. In Figs.~\ref{fig_poincare_L} and \ref{fig_poincare_H}, we show the sections in $(y,p_y)$ defined by $x=0$, colour-coded by $\lambda_{\rm max}$ and $\mathcal{I}_{\rm N}$, respectively, for different energies.
Both indicators, despite the greater scatter of $\mathcal{I}_N$ at nearly equal $\lambda_{\rm max}$, reveal how for $E\gtrsim0.8$ the fraction of accessible phase-space is indeed occupied by a rapidly growing fraction of chaotic orbits though with a persistent measure of regular islands, even for the $E=0.16$, closer to the $E=1/6$ threshold case. 
\begin{figure}
        \centering
        \includegraphics[width=1\columnwidth]{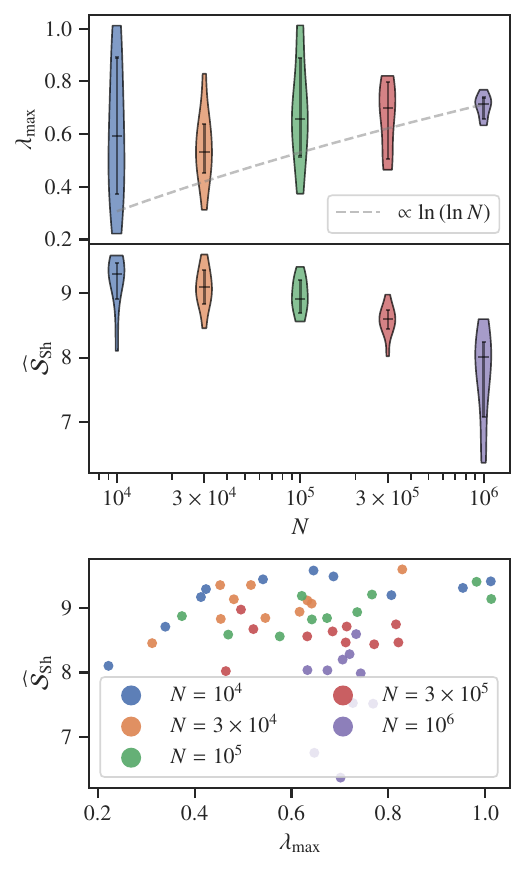}
        \caption{Top panels: Distributions of the largest Lyapunov exponent, $\lambda_\mathrm{max}$ (upper sub-panel), and Shannon entropy, $\widehat{\mathcal{S}}_{\rm Sh}$ (lower sub-panel), for different values of $N$ in the $N$-body system. Error bars indicate the median and the associated $1\sigma$ scatter. In the upper sub-panel, the dashed line denotes the scaling proposed by \citet{1993ApJ...415..715G}. Bottom panel: $\widehat{\mathcal{S}}_{\rm Sh}$ versus $\lambda_\mathrm{max}$ for individual realisations of the $N$-body system. The largest Lyapunov exponent shows no significant dependence on $N$, whereas the Shannon entropy decreases with increasing $N$.}
        \label{fig:lambdaevol}
\end{figure}
\subsection{$N$-body models}
We performed a series of $N$-dependent numerical experiments on test orbits for different values of the initial particle energy. As mentioned above, numerical studies on frozen $N$-body realisations of simple galactic potentials (see e.g. \citealt{2001PhRvE..64e6209K,2002PhRvE..65f6203S,2004PhRvS...7a4202K}) seem to suggest that individual particle trajectories, for increasing $N$ approach more and more closely their counterparts in the parent smooth system, while their largest Lyapunov exponent does not show a significant decreasing trend with $N$, as we would naively expect. 

In the range $10^4\leq N\leq 10^6$, we observe  a similar tendency even in our time-dependent $N-$body potentials for the orbits of test particles initialised with exactly the same phase space position approaches in realisations of the potential with increasing values of $N$.
As an example, in Fig.~\ref{fig:orbitnbody},  we show the trajectory for a particle of initial energy per unit mass $E\approx-0.66$ in Plummer models sampled by $N=10^4$, $3\times 10^4$, $10^5$, $3\times 10^5$, and $10^6$. Notably, as $N$ increases, the fluctuations of the orbital inclination become less and less important, being already negligible for $N=10^6$. A similar behaviour is observed for other values of the orbital eccentricity and initial energy (not shown here), with highly eccentric orbits at fixed $E$ (i.e. having lower values of angular momentum) being generally more subjected to wild fluctuations in the inclination, even at $N\approx 10^5$.

We note that $\lambda_{\rm max}$ evaluated as function of a given value of the energy $E$ depends weakly on $N$ (see also Fig.~2 in \citealt{2001PhRvE..64e6209K}, fig.~12 in \citealt{2019MNRAS.489.5876D} and Fig.~10 in \citealt{2025A&A...698A..28S}). This is evidenced in Fig. \ref{fig:lambdaN} where we tracked the evolution of the mean finite time largest Lyapunov exponent for the same orbital initial condition (i.e. test particle with equal combinations of $\mathbf{r}$ and $\mathbf{v}$), propagated in different realisations of the Plummer model at different values of $N$. Using the same number of independent samplings of the underlying model (shaded areas) the mean values of $\lambda_{\rm max}$ over the ensemble relax to comparable values over $10^3$ dynamical times, regardless of $N$.

We emphasise that unlike in the frozen-potential experiments of Kandrup and collaborators, the energy of the test particle is not conserved for any $N$, owing to the time-dependent potential generated by the $N$ moving field particles. It is therefore necessary to quantify the range of energies explored along each trajectory.

In the top panel of Fig.~\ref{fig:lambdae}, we show $\lambda_{\rm max}$ as a function of the median energy over the full integration. Orbits at lower energies (i.e. particles deep in the potential well) have larger values of their largest Lyapunov exponent, whereas the orbits exploring higher energies (i.e. weakly bound particles) systematically show smaller $\lambda_{\rm max}$. This trend is consistent with the relation between the averaged $\lambda_{\rm max}$ and the initial energy found in noisy-potential experiments mimicking an increasing number of field particles (up to $10^{13}$) by \cite{2025A&A...698A..28S}, as well as in low-resolution direct $N$-body simulations by \cite{2019MNRAS.489.5876D}. Furthermore, as $N$ increases, the scatter in $\lambda_{\rm max}$ about its relation with the median energy decreases. 

Because the particle energy is not conserved, we compared $\lambda_{\rm max}$ against both the initial and median energy at fixed $N$. As shown in the lower panels of Fig.~\ref{fig:lambdae}, both $\lambda_{\rm max}$ and $\widehat{\mathcal{S}}_{\rm Sh}$ decreases monotonically with both quantities. Notably, for weakly bound particles ($E\gtrsim-0.3$), the initial and median energies (green crosses) are nearly identical, as expected for orbits confined to the lower-density outskirts of the cluster.

With this established, the top and middle panels of Fig.~\ref{fig:lambdaevol} show the violin plots of $\lambda_{\rm max}$ and the information entropy, $\widehat{\mathcal{S}}_{\rm Sh}$, respectively, including their statistical spread, for orbits with initial energy $E=-0.45$ (corresponding to a circular orbit of radius $a$ in the smooth Plummer potential). The orbits are evolved across ensembles of discrete realisations with increasing $N$.

The distribution of the largest Lyapunov exponent shows no significant dependence on $N$ over this range, while its spread narrows markedly with increasing $N$. \REV{At first sight, this appears inconsistent with the $\ln(\ln N)$ scaling proposed by \citet{1993ApJ...415..715G}. However, their estimate was derived assuming the propagation and amplification of independent perturbations through strong two-body encounters in a fully self-gravitating $N$-body system. In contrast, our setup follows a test particle evolving in the live potential generated by $N$ particles and, therefore, it does not capture the same self-consistent amplification mechanism.} In contrast, the entropy decreases with $N$ and exhibits a broader spread, particularly at $N\sim10^6$. This trend is also evident in the bottom panel, which shows $\widehat{\mathcal{S}}_{\rm Sh}$ as a function of $\lambda_{\rm max}$ for individual realisations. At fixed $N$, larger Lyapunov exponents are broadly associated with higher entropy, although the correlation is weaker than it is for the HH system.

\section{Discussion and conclusions}\label{sec:conclusions}
We investigated the relation between the largest Lyapunov exponent and different formulations of information entropy in the paradigmatic \cite{henon1964} 2D potential and for individual particle orbits in time-dependent $N$-body potentials. Our main results can be summarised as follows.

For the HH model, the Shannon entropy averaged over trajectories on the same energy surface scales logarithmically with energy, with different pre-factors in the ${\lesssim}0.08$ and ${\gtrsim}0.08$ regimes (Fig.~\ref{fig:lmaxHH}). This change in the pre-factor reflects the transition from weakly to strongly chaotic motion across the well known onset of widespread chaos in the HH system. This behaviour is consistent with that of the largest Lyapunov exponent and of the mutual information entropy. The latter scales approximately linearly with $\lambda_{\rm max}$, whereas $\widehat{\mathcal{S}}_{\rm Sh}$ exhibits a logarithmic dependence (Fig.~\ref{fig:scatter2}).

For single-particle orbits in $N$-body realisations of a given gravitational potential (here, a spherical Plummer model), $\lambda_{\rm max}$ is larger for more tightly bound orbits (i.e. lower energies), while it decreases for less bound trajectories; the same qualitative trend is observed for the Shannon entropy (Fig.~\ref{fig:lambdae}). At fixed initial conditions, increasing the resolution of the $N$-body system (i.e. larger $N$ at fixed total mass) produces little variation in $\lambda_{\rm max}$, even as the dynamics approaches the continuum limit (Fig.~\ref{fig:lambdaevol}). In contrast, the Shannon entropy decreases monotonically with $N$, largely independently of $E$, in qualitative agreement with frequency-map analyses of frozen $N$-body systems \citep{2001PhRvE..64e6209K,2002PhRvE..65f6203S}. \REV{This discrepancy suggests that $\lambda_{\rm max}$ alone is not a sufficient indicator of chaoticity in gravitational $N$-body systems, at least if chaoticity is understood in terms of finite-time phase-space exploration. The maximum Lyapunov exponent measures the fastest local rate of infinitesimal stretching in tangent space. In contrast, $\widehat{\mathcal{S}}_{\rm Sh}$ measures the finite-resolution volume effectively sampled by an orbit over the integration time. These two
quantities are expected to correlate when local stretching efficiently produces transport across the accessible region of phase space and when the dominant stretching direction is representative of the overall instability. However, this connection need not hold in general. In this sense, the entropy indicator is not a replacement for the Lyapunov exponent, but a complementary diagnostic sensitive to the macroscopic region explored by the trajectory. Although $\widehat{\mathcal{S}}_{\rm Sh}$ is not an estimator of the KS entropy per se, it is motivated by the same connection between instability, information production, and phase-space exploration. Its behavior could therefore reflect dynamical information that is not contained in $\lambda_{\rm max}$ alone, but is instead associated with the collective contribution of the positive Lyapunov exponents.}

We conjecture that as $N$ increases, $\lambda_1\equiv\lambda_{\rm max}$ remains approximately constant (at least up to $N=10^6$), while the second and third exponents $\lambda_2$ and $\lambda_3$ decrease significantly beyond $N\sim10^5$. This interpretation is supported by Fig.~\ref{fig:orbitnbody} ($N=10^6$), where perturbations in the precession frequency persist, but fluctuations of the orbital plane are strongly suppressed, indicating a near conservation of the angular momentum. We note that in a related context, the inclusion of post-Newtonian corrections can decrease the leading exponent, while increasing subdominant ones in certain configurations of the three-body problem \citep{2025A&A...693A..53D}. 

\REV{If the decrease in $\widehat{\mathcal{S}}_{\rm Sh}$ with $N$ is driven by the suppression of sub-leading unstable directions, or more generally, by the progressive smoothing of the finite-$N$ potential, then we should expect $\widehat{\mathcal{S}}_{\rm Sh}$
to approach an asymptotic value at a sufficiently large $N$. This limiting value need not be connected to the saturated value of $\lambda_{\max}$. Thus $\lambda_{\max}$ may remain approximately constant, while $\widehat{\mathcal{S}}_{\rm Sh}$ decreases and eventually saturates to the value associated with the corresponding smooth-potential orbit, for the adopted integration
time and coarse graining. In a smooth spherical Plummer potential, this baseline is expected to reflect regular motion constrained by the integrals of the mean-field problem, rather than continued chaotic exploration.}

A direct test would require computing the full Lyapunov spectrum (Eq.~\ref{eq:spectrum1}) for the test particle. However, this entails orthonormalisation in a $6N$-dimensional phase space, which requires evolving and orthonormalising a $6N\times6N$ tangent basis and becomes computationally prohibitive already at $N\approx10^4$. An alternative is to track the deformation of an initially hyperspherical ensemble of trajectories, as proposed by \citet{2025A&A...698A..28S}.

In light of these results, we speculate that the $N$-body chaos driven by close encounters \citep[sometimes referred to as punctuated chaos, see][]{2023MNRAS.526.5791P,2023IJMPD..3242003B}, which appears to increase with $N$ \citep{2023AIPC.2872e0003P}, primarily reflects the behaviour of the largest Lyapunov exponent, which probes local instability driven by close encounters, rather than global phase-space transport. In contrast, a dynamical entropy measure, being sensitive to global mixing through the full Lyapunov spectrum, would decrease with $N$, owing to the suppression of subdominant exponents. We emphasise that trajectory-based entropy measures provide a viable alternative to $\lambda_{\rm max}$ only in low-dimensional systems (e.g. driven 1D models or 2D non-integrable potentials such as the HH system), where the sum of the positive part of the Lyapunov spectrum closely traces $\lambda_{\rm max}$. In higher dimensional phase spaces, quantities such as $\widehat{\mathcal{S}}_{\rm Sh}$ can instead serve as diagnostics indicating whether a full spectral analysis is required.

Finally, alternative estimates of dynamical entropy can be obtained from the compressibility of time series (e.g. from trajectory data), for instance via Lempel-Ziv-type complexity measures \citep{lempelziv1976,2003PhyD..180...92P}. A natural extension of this work is to compare, for fixed system parameters (e.g. energy), trajectory-based entropy estimates with those derived from compression schemes. This approach may be particularly useful when variational equations are unavailable or computationally expensive or when dealing with observational data (e.g. minor-body ephemerides or Galactic-centre S-stars), where only finite time series are accessible.
 
\begin{acknowledgements}
We thank the anonymous referee for their insightful comments, which have improved the manuscript. PFDC acknowledges the support from the MUR PRIN2022 project “Breakdown of ergodicity in classical and quantum many-body systems” (BECQuMB) Grant No. 20222BHC9Z. AAT and PFDC are grateful to Anna Lisa Varri and the University of Edinburgh for their hospitality during the 2nd ``Chaotic rendezvous'' meeting, and to the IFPU for hosting the workshop ``Chaos and nonlinearity in dynamical astronomy''. AAT dedicates this work to the memory of his father, Vincenzo Trani, a mathematics teacher who inspired his interest in science. This study was completed during a period of personal recovery, and he gratefully acknowledges the support received during this time.
\end{acknowledgements}
\bibliographystyle{aa} 
\bibliography{totalms} 
\end{document}